\documentclass[aps,pre,preprint]{revtex4}
\usepackage{amsmath,bm,epsfig}
\let \nn  \nonumber

\let\*\cdot
\def\<{\left\langle} \def\>{\right\rangle} \def\({\left(} \def\){\right)}
\let\p\partial \let\~\widetilde \let\^\widehat 

\def\be{\begin{equation}}\def\ee{\end{equation}}
\def\bea{\begin{eqnarray}}\def\eea{\end{eqnarray}}
\def\bse{\begin{subequations}}\def\ese{\end{subequations}}
\newcommand{\BE}[1]{\begin{equation}\label{#1}}
\newcommand{\BEA}[1]{\begin{eqnarray}\label{#1}}
\newcommand{\BSE}[1]{\begin{subequations}\label{#1}}

\let \nn  \nonumber
\usepackage{pstricks}
\usepackage{pst-node}
\usepackage[ansinew]{inputenc}
\usepackage{amssymb,amsmath}

\def\BSE{\begin{subequations}}\def\ESE{\end{subequations}}
\let \= \equiv

\def\p{\partial}

\def\b{\beta}

\def\o{\omega}

\def\be{\begin{equation}}       \def\ba{\begin{array}}

\def\ee{\end{equation}}         \def\ea{\end{array}}

\def\bea {\begin{eqnarray}}      \def\eea {\end{eqnarray}}

\def\bean{\begin{eqnarray*}}    \def\eean{\end{eqnarray*}}

\def\e{\varepsilon}

\def\<{\langle} \def\({\left(}  \def\>{\rangle} \def\){\right)}

\newtheorem{exi}{Example}

\begin{document}

\title{Discrete exact and quasi-resonances of Rossby/drift waves on $\b$-plane with  periodic boundary conditions}
\author{A. Kartashov$^\sharp$, E. Kartashova}
 \email{alexkart1@gmx.at, Elena.Kartaschova@jku.at}
  \affiliation{$^\sharp$AMS, Waidhofen an Ibbs, Austria}
  \affiliation{$^*$Johannes Kepler University, Linz, Austria}
   \begin{abstract}
Analysis of resonance clustering in weakly nonlinear dispersive wave systems, also called discrete wave turbulent systems, is a new methodology successfully used in the last years for characterizing energy transport due to exact and quasi-resonances. Quite recently this methodology has been used in the paper by M. D. Bustamante, U. Hayat "Complete classification of discrete resonant Rossby/drift wave
triads on periodic domains", \cite{BH13}, in order to show that resonance clustering is very sparse and quasi-resonances (that is, resonances with small enough detuning) play  major role in the energy transport in this specific wave system. On the other hand, in the paper by M. Yamada, T. Yoneda "Resonant interaction of Rossby waves in two-dimensional flow on $\beta$-plane", \cite{YaYo13}, the same physical system is studied and a mathematically rigorous theorem is proven: at high $\b$, the flow dynamics is governed exclusively by resonant interactions.  In our present paper we demonstrate that this seeming contradiction between numerical results \cite{BH13} and analytical results \cite{YaYo13} are due to some pitfalls in numerical studies of exact and quasi-resonances presented in \cite{BH13}. We also demonstrate that resonance clustering of drift waves on periodic $\b$-plane differs substantially from characteristic resonance clustering in other 3-wave systems: instead of a usual set of isolated triads and a few bigger clusters, there exists \emph{no isolated triads} in this case. Resonant triads are interconnected in a complicated way and  the smallest cluster consists of 6 connected triads and is formed by 6 distinct modes only. This specific clustering is due to the form of the dispersion function which allows in the general case 12 symmetries; this yields a special mechanism of energy confinement by exact resonant modes which is lacking in other presently known three wave systems and even in the system of drift waves with other boundary conditions.
\end{abstract}


\maketitle
\section{Introduction}\label{s:int}
Importance of studying resonances and quasi-resonances in weakly nonlinear dispersive wave systems  is due to the fact that, if nonlinearity is small enough, they describe complete dynamics of a wave system. Mathematical formalism in this case can be presented  as follows (in this paper we restrict ourselves to  three wave systems, for more details and four wave systems see \cite{CUP}).

Regard a wave system  governed by an evolutionary nonlinear partial differential equation with a small nonlinearity
\be \label{in-eq}
L(\varphi)=-\e N(\varphi)
\ee
where $L$ is a linear operator such that
\be \label{lin-eq}
L(\psi)=0, \, \psi=A \exp{[i(\mathbf{x}\cdot \mathbf{k} - \o \cdot t)]}, \, A=const,
\ee
$N$ is an arbitrary nonlinearity and $0<\e\ll 1$ is a small parameter. Then three wave resonance conditions have the form
\be\label{res-con}
\o_1 \pm \o_2 \ \pm \o_3=0, \, \mathbf{k}_1 \pm \mathbf{k}_2  \pm \mathbf{k}_3=0,
\ee
and amplitudes of the resonantly interacting waves $A_j, \, j=1,2,3, $ are not constant anymore but are slowly changing functions of a new variable - the so-called slow time $T=t/\e$.  Dispersion relation $\o=\o(\mathbf{k})$ is defined by (\ref{lin-eq}) and the form of boundary conditions.

The use of any multi-scale method, \cite{Na81}, allows us to deduce dynamical equations for amplitudes $A_j(T)$ of three resonantly interacting waves which have a simple  form in canonical variables $B_j(T)$
 \bea \label{complexA}
\dot{B}_1=  V_{12}^{3} B_3B_2^* , \ \
\dot{B}_2= V_{12}^{3} B_1^*B_3,  \ \
\dot{B}_3=  -V_{12}^{3} B_1B_2,
 \eea
for an arbitrary three wave system encountered in fluid mechanics, geophysics, plasma physics, astronomy, etc. Here the notation $\dot{B}_j$ is used for $d B_j/d T.$

The difference between all these physically different phenomena is hidden in the form of the interaction coefficient $V_{12}^{3}$  depending on the
initial equation (\ref{in-eq}), chosen solution of resonance conditions (\ref{res-con}) and  form of dispersion function $\o(\mathbf{k})$. If the interaction coefficient ivanishes, $V_{12}^{3} = 0$ for a chosen solution of (\ref{res-con}), this solution is dynamically not relevant and can be omitted. Accordingly, resonance conditions should be supplemented by the condition of non-vanishing interaction coefficient,
\be \label{in-coef}
V_{12}^{3} \neq 0.
\ee
The choice of canonical variables $B_j(T)$ allows to characterize dynamics in terms of one coefficient $V_{12}^3$ but the $B_j(T)$ are some functions of physical wave amplitudes $A_j$. Dynamical system on amplitudes $A_j$ has three different interaction coefficients satisfying some additional condition (examples are given in \cite{CUP}).

Study of nonlinear PDEs with periodic boundary conditions makes it necessary to look for the solutions of
(\ref{res-con}),(\ref{in-coef}) in integers. Since centuries the search for integer or rational points of manifolds or elliptic curves  was and still is an open mathematical problem. Accordingly in early 1960s two mathematical objects have been mainly studied in weakly nonlinear wave systems: (\textbf{a}) an isolated resonant triad, and (\textbf{b}) an infinite number of interconnected resonant triads. The approach (\textbf{b}) allows to smooth out dynamics of distinct modes and to obtain -  instead instead of an infinite number interconnected dynamical systems -  one wave kinetic equation at the time scale $t/\e^2$; the kinetic equation  is studied in the frame of kinetic wave turbulence theory, \cite{ZLF92},
 which is outside  the scope of present paper.

The  approach(\textbf{a}) allows to describe dynamics of distinct resonantly interacting modes on time scale $t/\e$. In this case, for
a chosen wave vector  its resonance curve has been usually constructed which is the locus of pairs of wavevectors interacting resonantly  with a given wavevector $\mathbf{k}_o$. For instance, resonance curves  for two dimensional drift waves on the $\b$-plane can be found in \cite{LongHigGill67} where it has been shown that depending on the choice of $\mathbf{k}_o$ the locus might be an ellipse, might be shaped like an hour-glass or  degenerate into two intersecting lines. However, there is no general method for finding integer points of a resonance curve or manifold (\ref{res-con}), or at least to establish whether they exist. Example is  given in \cite{PHD1} of a three wave system - capillary water waves with periodic boundary conditions - where  exact resonances are absent. To prove this fact analytically, the frequency resonance condition for this case has been reduced to the Last Fermat theorem $x^p+y^p=z^p$ with $p=3$ which demonstrates variety of problems which might be accounted for while looking for integer points on resonant manifolds.

A novel method of representing the complete set of resonant triads in a finite spectral domain as a set of resonant clusters consisting of one isolated triad or a few triads having common modes has been developed and applied in  early 1990s, \cite{KPR,PHD1,PHD2,PRL}, though the  notion of "resonance cluster" has been first introduced in 2007, \cite{KL07}. Complete methodology used to compute a set of resonance clusters, the corresponding set of dynamical systems and to describe energy percolation in the Fourier space via resonance triads can be found in \cite{CUP}.
 This methodology has been successfully used to clarify numerous results of numerical simulations and laboratory experiments, and also  to model real physical phenomenon - intra-seasonal oscillations in the Earth atmosphere,  \cite{KL07}.

In the present paper we apply these methods to construct and to study resonance clustering based on data kindly provided to us by M. D. Bustamante and U. Hayat, \cite{BH13}, and by M. Yamada and T. Yoneda, \cite{YaYo13}, and results of our own simulations.

Our paper has the following structure. In Sec.\ref{s:ex} we give a brief overview of results on resonance clustering available for the Charney-Hasegawa-Mima equation (CHME) in various bounded domains.  In particular we demonstrate that "complete classification of resonant triads" constructed in \cite{BH13} is not only incomplete  but in fact misleading. Indeed, it can easily be shown analytically that the smallest resonance cluster in the CHME on the $\b$-plane with periodic boundary conditions is \emph{not an isolated triad} as it is stated in \cite{BH13} but a cluster of 6 connected triads which suggests completely different physical implications.

   In Sec.\ref{s:quasi} we explicate the construction of quasi-resonant triads given in \cite{BH13} and show why this is grossly incomplete and statistically biased.

In Sec.\ref{s:altappr} we propose another approach to the problem of constructing resonance clustering and give an outline of our algorithm, based on the notion of resonant curves, introduced for this case as early as in 1967 in \cite{LongHigGill67}. This algorithm has cubic computational complexity, $T=\mathcal{O}(L^3)$, which results in quite realistic computation time (according to preliminary tests on a computer comparable to one used in \cite{BH13} estimated as a few hours to a few days, depending on implementation details) and has the evident advantage of finding \emph{all} exact solutions of the three wave kinematic resonant conditions.
We also show how a modification of this algorithm can be used to find all quasi-resonant solutions for a given detuning level.  Brief discussion in Sec.\ref{s:diss} concludes the paper.

\section{Resonant triads}\label{s:ex}

The CHME on a sphere with rigid lid (also called barotropic vorticity equation) describes large-scale processes in the Earth's atmosphere and has been studied by many researchers starting with \cite{Char47}; other classical geophysical references can be found in the monograph \cite{ped}. Hamiltonian formalism for atmospheric Rossby waves has been first presented in \cite{Pit98}.

In the context of resonance clustering  a few  main points of reference on spherical drift/Rossby waves are: \cite{sil} (conditions for non-vanishing of the interaction coefficient),  \cite{KPR} (the first resonance clustering found in the spectral domain $|m|,|n| \le 50$), \cite{KK06-3} (a fast numerical algorithm which allows to compute exact resonances in domains of $|m|,|n| \sim 10^{6}$), \cite{LPPR09} (study of energy percolation in the domain $|m|,|n| \le10^{6}$), \cite{KL07} (a phenomenon of intra-seasonal oscillations in the Earth atmosphere is interpreted \emph{via} dynamics of isolated resonant triads of spherical Rossby modes).
Dispersion function in this case has the form
\be \label{dis-atm}
\o(m,n)_{atm}=-2m/[n(n+1)].\ee

The CHME on the $\b$-plane used in geophysics to describe large-scale quasi-geostrophic motions in a barotropic rectangular ocean $[0, L_x] \times [0,L_y]$ of constant depth with a rigid lid. In this case the CHME takes the form
 \be \label{CHM}
\frac{\p}{\p t} \triangledown^2 \psi + J(\psi,\triangledown^2 \psi) + \beta  \frac{\p \psi}{\p x}=0,
\ee
and should be solved with non-flow boundary conditions: $\psi=0$ for $x=0, L_x$ and $y=0,L_y.$  Here $\b$ is the derivative of the Coriolis parameter with respect to latitude, in geophysics $\b \approx 0.127.$ Computation of nteraction coefficients, an algorithm for computing resonance clustering in large spectral domains, and examples of clustering are given in \cite{KR}, \cite{KK06-1} and  \cite{all08} correspondingly.
 Dispersion function in this case has the form
\be \label{dis-ocean}
\o(m,n)_{ocean}=-\b /[2\pi \sqrt{m^2+n^2}].\ee

In both cases - the CHME on a sphere and on the $\b$-plane with zero boundary conditions - the characteristic properties of resonance clustering are similar to every other three wave system studied before: not more than $20 \div 30 \%\% $ of all modes participate in exact resonances and most resonant clusters are isolated triads.  As dynamical system for an isolated triad is integrable, this means that at the time scale $t/\varepsilon$ corresponding resonant modes evolve periodically while non-resonant modes just keep their energy (see e.g. Fig.1 and Fig.2, \cite{PRL}, spherical modes in full numerical spectral model of the CHME).

The CHME with periodic boundary conditions on the $\b$-plane is one of the most famous equations in plasma physics describing drift waves in tokamak. It was first written out in the pioneering paper of Hasegawa and Mima, \cite{HM78}, in 1978, and later generalized by Wacatani and Hasegawa in \cite{WH84}. Further developments are recently reviewed in \cite{DIIH05,DHM11} and in monograph \cite{DII10}.

The resonance clustering  presented in \cite{BH13} for this case and dispersion function of the form
\be \label{dis-plasma}
\o(m,n)_{plasma}=-\b m /(m^2+n^2)
 \ee
has the same characteristic structure as in the geophysical examples given above. However, for this specific case it is rigorously proven that at high $\b$, flow dynamics is governed exclusively by resonant interactions, \cite{YaYo13}. This result suggests that resonance clustering can not be so sparse as in the other three wave system, and this was our motivation to study more closely the method of computing resonances is given in \cite{BH13}.

Finding (exact) resonant triads (which are described by a Diophantine equation) has been reduced in \cite{BH13} to finding rational points on a certain class of elliptic curves.
Elliptic curves and, specifically, their rational points, have at least for a century been one of the favorite objects of study in algebra and number theory, and a vast literature on them exists.
However, finding \emph{all rational points} of an elliptic curve is still an open question and the results of the paper fall short of the claim of "complete classification" suggested by the title. To estimate the number of omitted triads in the resonance set computed by this method, we used the complete set of resonant modes in the spectral domain $|m|,|n| \le 1000$, kindly provided by Michio Yamada and Tsuyoshi Yoned,  \cite{YaYo13}.

The set includes all the modes with both coordinates
within square box of size $L=1000$, i.e. $|m|, |n| \le 1000$ which
participate in {\emph some} resonance triad (their triad partners
probably lying outside the $L$-box). As we are interested in the study of resonant triads, the first task was to convert the list of resonant modes into a list of resonant triads, all three modes of which lie within a given $L$-box.
In the present study we limit ourselves to the domain $L=200$.
This is enough to draw well-founded conclusions on the frequency of
resonance occurrence and structure of more common resonance clusters;
we also wanted to enable comparability to the results of \cite{BH13}, where the
same problem has been dealt with by other methods.

Having a list of all resonant modes in some domain, the method of
reconstruction triads in that domain is quite straightforward, though some implementation details are a little tricky.
Truncation of modes' list to $L=200$ is trivial.
Now consider the set of all pairs of modes from the truncated list and treat them as the
first two modes ${m_1, n_1}, {n_2, m_2}$ of a (possible) resonant triad. The third mode of each triad is uniquely constructed via the linear conditions, i.e. $m_3=m_1+m_2, n_3=n_1+n_2$.
First we discard those triads whose third mode lies outside of our box, $|m_3| \ge 200$ or $|n_3| \ge 200$.
For all other triads the resonance condition should be checked. This can be computed either in integers or in floating-point numbers (called "real" everywhere below).
Calculation in integers always gives the precise answer, while dealing with real numbers can produce a small computational error.
However the problem with integers is, that direct calculation leads to dealing with numbers up to the order of $3 \times 10^{10}$, which lies outside the "long integer" range of usual computers.

This being the case, we chose the following reasonable compromise.
We compute the resonance condition $\omega_1 + \omega_2 - \omega_3 = 0$ in reals. For most resonant triples it holds exactly, while some show a small deviation (computation error artefact ). Also for some non-resonant triples this condition holds up to a very small deviation (quasi-resonant triads). Now we order the list of triads by their deviation ascending and truncate the list at some deviation $\epsilon$ large enough not to discard any resonant triad, say $\epsilon= 10^{-8}$. Triads with zero deviation are
immediately placed into resonant triads' list, while the rest must be checked precisely, i.e. in integers.
We have developed techniques for dealing with "oversized" integers before, \cite{KK06-3}, and it can be used here with some minor alterations.

This method is, however, rather cumbersome, involving computation of quite a few GCDs (greatest common divisors) of large numbers so for efficiency reasons we did not apply it to the whole set of candidate triads.

We obtain a set of $N=828$ (\emph{versus} 133 shown in Fig.2, \cite{BH13}, for the same spectral domain $|m|,|n| \le 200$) resonant triads and our next goal is to study its structure. (Here and below  all results are given for triads with non-zero interaction coefficient.)
The bulk data is shown in the Table \ref{t:clusters} and it can be seen immediately that unlike the examples discussed above, resonant triads in this case are densely entangled: there are \emph{no isolated triads at all}. The total number of clusters is 72, including  40 clusters consisting of 6 interconnected triads.
\begin{table}
\begin{tabular}{|l||l|l|l|l|l|}
\hline
Number of triads in a cluster &6& 12 & 18 & 24 & 30\\
\hline
Number of clusters & 40 & 12 &12 & 2 & 6 \\
\hline
\end{tabular}
\caption{ \label{t:clusters} Overall data on the form and number of resonance clusters in the domain $|m|,|n|\le 200.$}
\end{table}

\begin{figure}
\includegraphics[width=14cm]{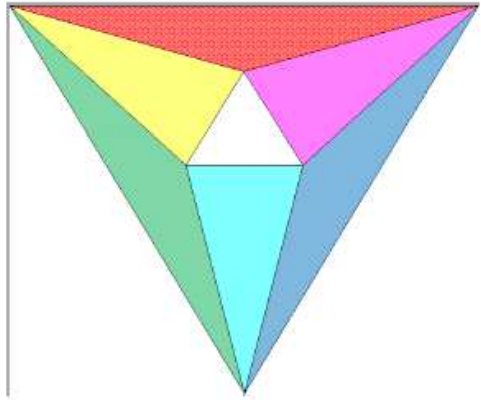}
\caption{\label{f:6tr} Color online. Topological structure of a 6-triad cluster (\ref{ex-6tr}); resonant triads are shown as colored triangles, the  white triangle in the middle  does not represent a resonant triad.}
\end{figure}

An example of a cluster of six triads which is \emph{the smallest cluster size} in the whole set is given below:
\bea \label{ex-6tr}
\begin{cases}
(-18,	46),(	2,	22),	(-16,	68),\\
(-18,	46),(	16,	-68),(	-2,	-22),\\
(-16,	68),	(-2,	-22),(	-18,	46),\\
(-16,	68),	(18,	-46),(	2,	22),\\
(-2,	-22),	(18,	-46),(	16,	-68),\\
(2,	22),(	16,	-68),(	18,	-46).\\
\end{cases}
\eea
Each triad of this cluster is connected to four others, having one common mode with two triads and
two common modes (an edge) with two others. The standard (plane) topological representation of this cluster is shown in  Fig.\ref{f:6tr}; it allows to  reconstruct the corresponding dynamical system, \cite{KM07}, and to characterize the type of energy percolation within the cluster, \cite{KL08}.

It can be checked by direct substitution into resonance conditions (\ref{res-con}) with dispersion function (\ref{dis-plasma}) that

\emph{Any resonant triad $(m_1,n_1),(m_2,n_2),(m_3,n_3)$ participates in a cluster of 6 connected triads.
There also exists a conjugate cluster of 6 triads, generically not connected with the first
but consisting of modes with same coordinates, up to some sign changes.}

In particular, the 6-triad cluster which is conjugate to (\ref{ex-6tr}) consists of:
\bea \label{ex-6tr-conj}
\begin{cases}
(-18,-46),(2,-22),(-16,-68), \\
(-18,-46),(16,68),(-2,22),\\
(-16,	-68),(	-2,	22),(	-18,	-46),\\
(-16,	-68),	(18,	46),(	2,	-22),\\
(-2,	22),(	18,	46),(	16,	68),\\
(2,	-22),(	16,	68),(	18,	46).\\
\end{cases}
\eea

This means that a sixtuple of triads - and not an isolated triad -  is the primary building block of the cluster structure for drift/Rossby waves with periodic nonzero boundary conditions on the $\b$-plane.

Symmetries demonstarated above are best understood in terms of group action on a set.
Our triads lie in the $\mathbb{Z}^6=\{m_1, n_1, m_2, n_2, m_3, n_3\}$ space on the 4-dimensional hyperplane defined by linear conditions
\be \label{res-lin}
\mathbf{k}_1 \pm \mathbf{k}_2  \pm \mathbf{k}_3=0.
\ee
On this hyperplane a finite symmetry group of order 12 maps every generic
triad on twelve triads with equal detuning $\delta$. In particular, a triad with
detuning zero (resonant) is mapped on twelve resonant triads which are combined
into two clusters of six triads each.

\section{Quasi-resonant triads}\label{s:quasi}
Conditions for a quasi-resonance read
\be\label{res-con-quasi}
|\o_1 \pm \o_2 \ \pm \o_3|=\delta, \, \mathbf{k}_1 \pm \mathbf{k}_2  \pm \mathbf{k}_3=0,
\ee
with some small $0<\delta \ll 1$ which is called resonance width or frequency mismatch  or detuning etc.

Quasi-resonances take place at some later time-scale than exact resonances and if they are relevant for the dynamics of a specific wave system depends on the system. As it was mentioned before, in the CHM equation with big enough $\b$ quasi-resonances can be disregarded. On the other hand, in \cite{9} Rossby waves with forcing are regarded and it is shown that  for moderate Rossby number $\b\approx 0.1$, a reduced numerical model including near-resonances enables efficient energy transfer from three-dimensional forced modes to two-dimensional large-scale modes. Another example can be found
in \cite{10} where three classes of resonances between surface and interfacial waves are regarded in a two-layer density-stratified fluid. It is shown that resonances of the class III (two surface waves and
one interfacial wave travel in the same direction, and the interfacial wave has typically
a much longer wavelength compared to the two surface waves) "undergoes a cascade of (near-)resonance interaction that spreads the energy
of initial waves to a number of lower and higher harmonics". Resonance conditions in this case are written in terms of two different dispersion functions - one for surface waves, $\o_S$, and one for interfacial waves, $\o_I$ (see also resonant curves in Fig.1, \cite{10}). The importance of quasi-resonances is shown then analytically and numerically by studying the corresponding \emph{dynamical equations}.

Not going into more detail about the role of quasi-resonances in three wave systems, we would like to stress once more a very important point: whether quasi-resonances are important in some specific physical problem, \emph{can not be decided} from pure kinematic considerations or by analogy. This problem can only be investigated in the frame of reduced dynamical models, e.g. \cite{9,10}, or by studying an initial NPDE, e.g. \cite{YaYo13}.

As
 in some problem  settings depending on the choices of NPDE(s),  boundary conditions, number of space dimensions, choice of parameters, etc.,   it may happen that quasi-resonances'  play an important part in the energy transport in a three-wave system,  and a good numerical algorithm for studying quasi-resonances would be welcome.

However, the algorithm presented in \cite{BH13} for computing quasi-resonances will not generate a representative set of quasi-resonances for a very simple reason. This algorithm is looking for quasi-resonances which are formed in the $\delta$-vicinity of \emph{an exact resonant triad}, while  conditions (\ref{res-con-quasi}) are more general and can be satisfied by three wave vectors such that \emph{none of them is a resonant mode} and the triad lies far away from \emph{any resonant triad}. More precisely, we
introduce distance between two triads in the usual way, i.e. as Euclidean distance between 6-vectors:
\bea \label{dist-quasi}
D([(m_1,n_1)(m_2,n_2)(m_3,n_3)],[(m_1',n_1')(m_2',n_2')(m_3',n_3')]) =\nn \\
\sqrt{(m_1-m_1')^2+(n_1-n_1')^2+...+(n_3-n_3')^2}.
\eea

It is evident that for adjacent triads $T_1, T_2$, i.e. such triads, whose corresponding nodes lie on the same or adjacent nodes of the $2D$ integer grid $\mathbf{Z}^2$, $D(T_1, T_2) \le \sqrt{6}$.
Let us compare short-range and long-range search of quasi-resonances.

\begin{figure}
\includegraphics[width=16cm,height=12cm]{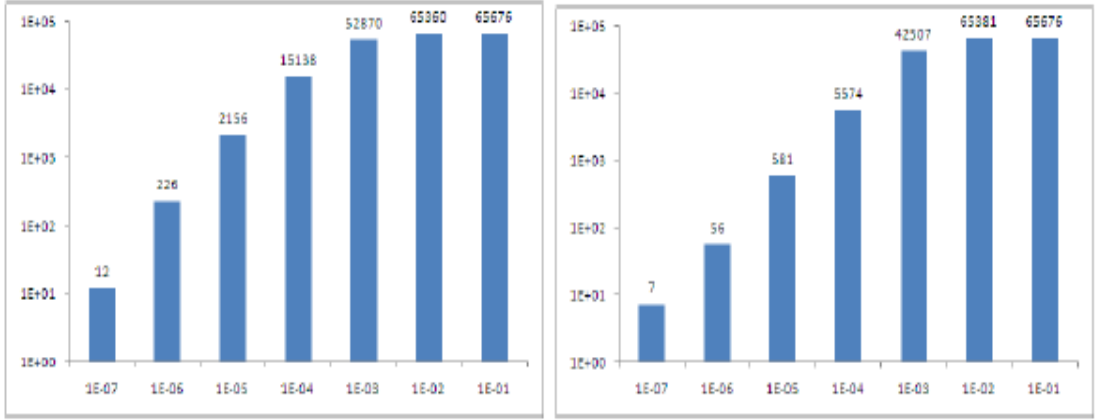}
\caption{\label{f:hist1} Color online.
Histograms of quasi-resonances, number of triads for detuning order, for triads in the nearest vicinity of exact resonant triads (left panel) and for the same number of randomly chosen triads (right panel)}
\end{figure}
\begin{figure}
\includegraphics[width=9cm,height=6cm]{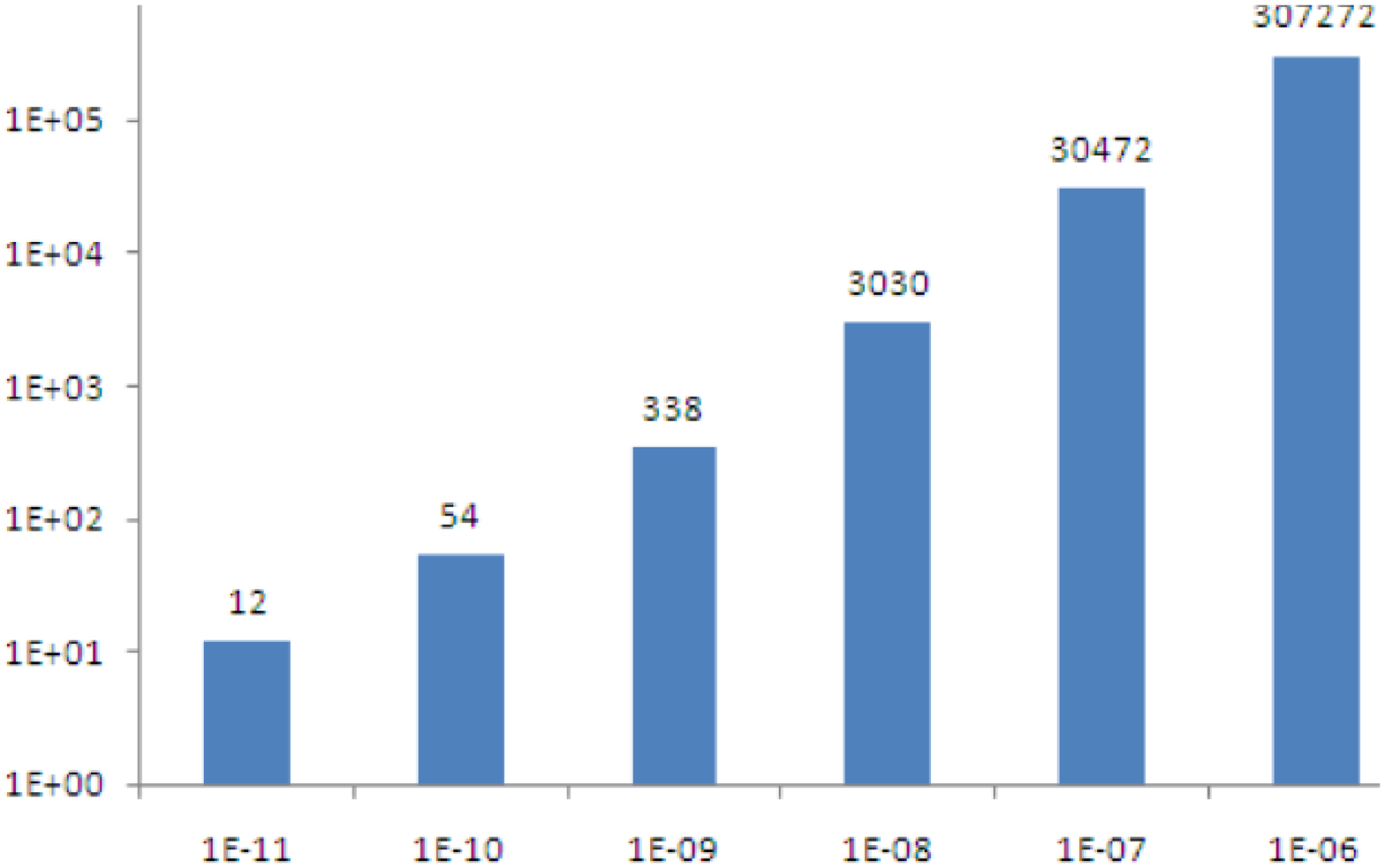}
\includegraphics[width=7cm]{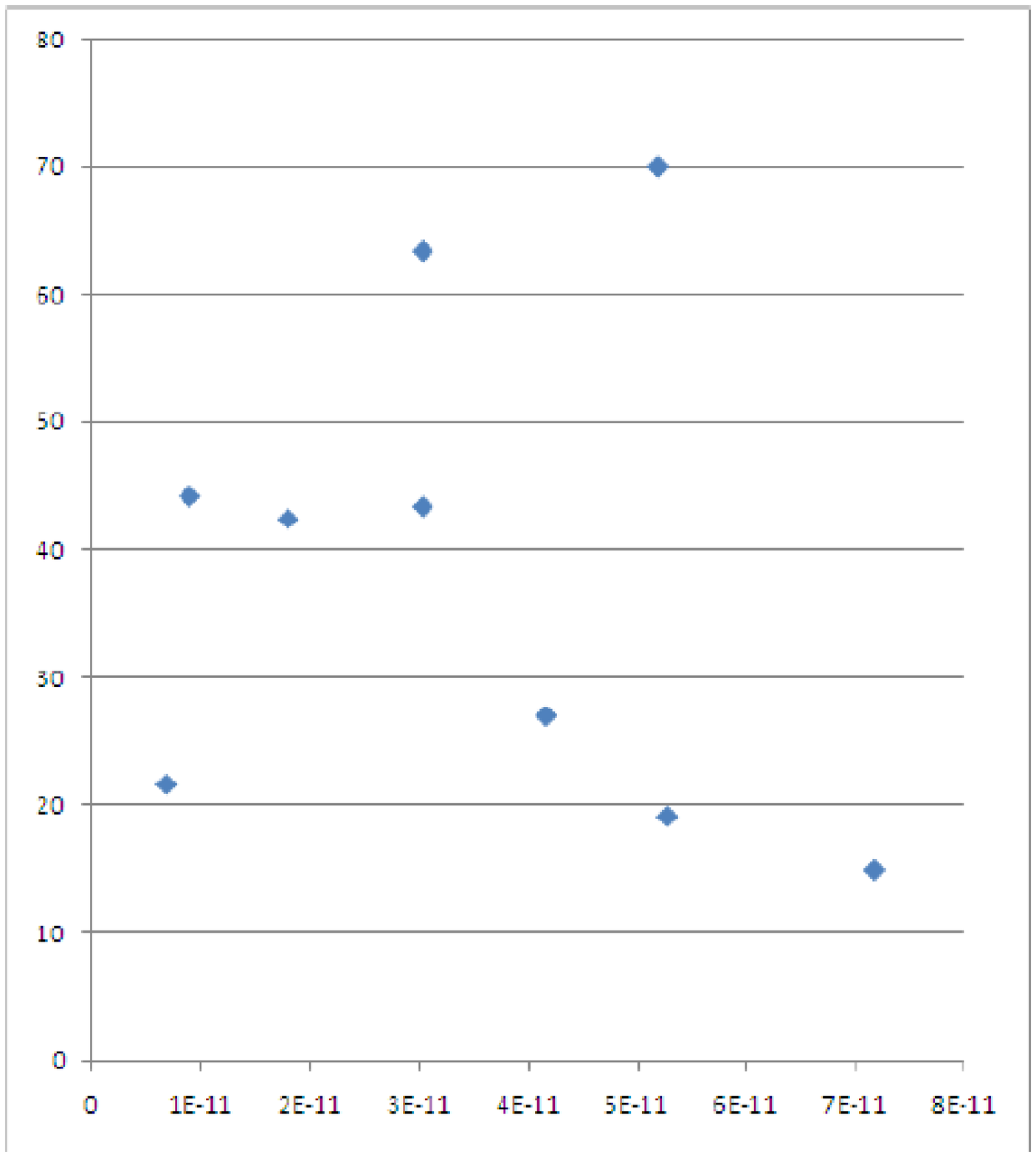}
\caption{\label{f:hist2} Color online. \textbf{Left panel}: Histogram for quasi-resonances found via full search in the domain $L=200$. \textbf{Right panel}: Quasi-resonant triads with very small detunings (horizontal axis), and their Euclidean distance (vertical axis) from the nearest exactly resonant triad.}
\end{figure}
For short-range search, we take each exact resonance triad and vary the coordinates of its three modes, either leaving a coordinate as it is or changing it by $\pm 1$. This (taking into account linear conditions) gives us up to $20$ adjacent near-resonant triads for each exact resonance triad (a few less for triads with coordinates on the border of the $L$-box). Altogether $ 65 576$ triads have been tested. The smallest detuning value found is
$\varepsilon_s \approx 6\cdot 10^{-7}$ and histogram of detuning distribution is shown in Fig.2 (left panel).
For each triad $T$ of this set the distance to the nearest resonant triad $D_{res}(T)$ is $\le \sqrt{6} \sim 2.45$.

We perform long-range search taking two first points at random and constructing the third according to linear conditions. The triad is accepted in the test pool if the third point lies within the $L$-box and non of the six coordinates is equal to zero.

We take as many random triads as have been tested for the short-range search (this number is fixed). Results of a typical run (these vary slightly from run to run, of course, due to random choice of triads) are presented in the same figure (right panel).

We immediately see that seeking for quasi-resonances in the vicinity of exact resonances has hardly any advantage over random choice.

Now to the full search. Testing \emph{all} triads in the domain $L=200$, we find much better quasi-resonances (triads with much smaller detunings) than either through short-range or random search. Their histogram is shown in the next figure (left pane).
A striking peculiarity of these quasi-resonances is, that none of them lie anywhere near a precise resonant triad. The scatter graph of distances to the nearest exact triad (in the sense of (\ref{dist-quasi}) against detunings (same figure, right panel) shows that triads with smallest detunings lie no nearer to precise resonances than $D=15$, and sometimes over $70$ (each point of the scatter graph corresponds to multiple triads due to symmetries discussed above).

One may argue that the search for quasi-resonances in some small vicinity of an exact resonance  might work for a special initial distribution of the energy in the wave field - energy should be  initially distributed only among exact resonant modes and at least in some triad(s) high-frequency modes have to be excited, but this should be explicitly discussed as it is done e.g. in \cite{PRL,KL08}.

However, even under suitable assumptions about initial energy distribution, the algorithm
 for seeking quasi-resonances presented in \cite{BH13} has a few critical flaws demonstrated and discussed below.

The construction starts with some irreducible (exact) resonant triad $(m_1,m_1),(m_2, n_2)(m_3,n_3)$ having at least one wavevector outside the $L$-box, $|m_i| > L$ or $|m_i| > L$, say the greatest wave number $m_1 = cL$, $c>1$ and rational. This triad is then scaled to fit into the box. In plain words, we consider the triad
\be
(m_1/c,n_1/c)(m_2/c, n_2/c)(m_3/c,n_3/c)
\ee
for which the resonance equations hold precisely, all wavenumbers lie within the $L$-box - but some of them are not integer any more.

Now consider all the triads with integer wavenumbers which lie in the $1$-vicinity of the scaled triad: we just round $m_i, l_i$ to the next integer value, up or down. In the general case we obtain $64$ triads, some of them probably sufficing the linear conditions $m_1+m_2=m_3, n_1+n_2=n_3$. From the latter we choose  the triad with the smallest detuning and call it quasi-resonant.

In this way, starting with a single irreducible triad outside the box, we construct {\it up to} $L$ quasi-resonant triads. (For some values of $L$ it can happen that none of the $64$ candidate triads suffice the linear constraints.) This process is repeated for every irreducible triad outside the box, and it is \emph{assumed} that
"\emph{the higher the box norm the smaller the size of the corresponding detuning level}". (p.2409, \cite{BH13}).

\begin{figure}
\includegraphics[width=16cm,height=8cm]{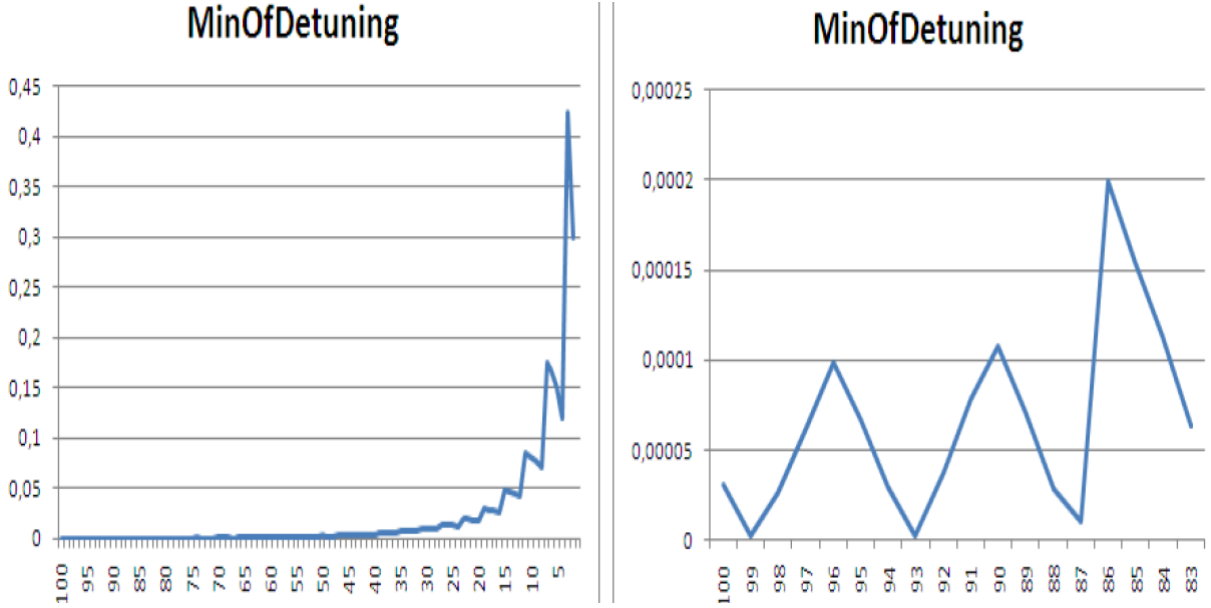}
\caption{\label{f:detun} Color online. Detuning as a function of the box-size $L$. For the triad $
[\{2, -49\}\{190, 155\}\{192, 106\}]$ detuning in shown for box sizes $L=100, 99... 1$ (on the left) and for $L=100, 99... 80$ (on the right)}
\end{figure}

This implies no less than that detuning value grows monotonously as the box size declines. However, this statement is not only unjustified, it is simply not true. Closely following the construction of the authors, we wrote a program which calculates detuning as a function of the shrinked box size and presents it graphically (available online, \cite{link}). Example of the detuning function is shown in the Fig.\ref{f:detun}. Indeed, we could not find a \emph{single} triad for which the assumption cited would hold.

The consequences of this error are near disastrous for the study of quasi-resonant triads presented in the \cite{BH13}. As an example, let us consider the triad
\be
(2, -49)(190, 155)(192, 106)
\ee
taken from Fig.1, \cite{BH13}. Choosing detuning level of $1.0 \times 10^{-4}$, we obtain quasi-resonant triads for box sizes $L=100, 99... 91$. For $L=90$ detunig is approximately $1.07 \times 10^{-4}$, so under the assumption of monotonous growth of detuning here we break off. But as detuning is not monotonous, there are four box sizes $L=89,88,87,83$ which fall into the chosen detuning range but are ignored by the construction proposed. This does not only reduce the number of near-resonant triads constructed but also introduces a systematical bias - triads with smaller wave numbers are ignored more often than those with larger ones - rendering further considerations on them (both statistical and structural) virtually worthless.
\begin{figure}
\includegraphics[width=16cm,height=8cm]{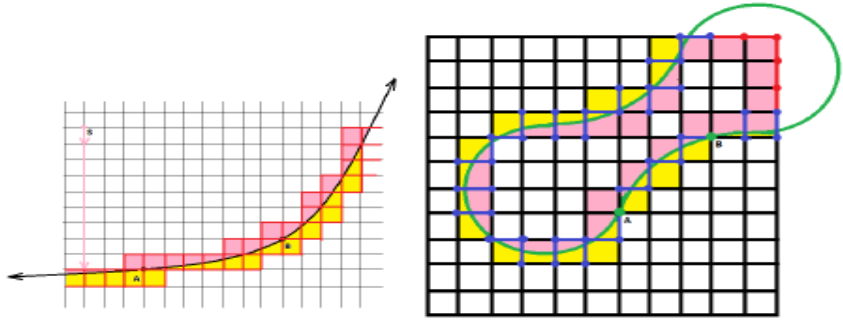}
\caption{\label{f:exact} Color online. Graphical presentation of using resonant curves
for fast computing exact and quasi-resonances (see Sec.IV).}
\end{figure}

\section{An alternative approach}\label{s:altappr}

For every mode ${m_1, n_1}$ within the box we should find all "partner" modes $(m_2, n_2)$ which would give rise to a resonant triad.
A brute force approach - testing all modes within the box - would give computational complexity $\mathcal{O}(L^4)$ which is unreasonably large.
However we make use of the fact that partner modes lie on a continuous smooth curve without self-intersections, called {\it resonance curve} and constructed  in \cite {LongHigGill67} (and some properties of such curves used below can be found there), and our task is reduced to finding integer points on this curve (if any). (For some special cases this curve degenerates into a couple of intersecting straight lines and finding integer points on them is quite easy.)

Step 1. Initial choice of $(m_1, n_1)$ and $(m_2, n_2)$.
We start with some triad $T=[(m_1,n_1)(m_2, n_2)(m_3,n_3)]$ within the box which gives an exact resonance. We insert the triad as the first member of the list of solutions. If we regard the wavevector ${m_1,n_1}$ as fixed and ${k_2, l_2}$ as variable \emph{with} $(m_2, n_2)$ \emph{real} and within the box ($|m_2|,|n_2| \le L$), exact resonances will lie on the resonance curve of $(m_1,n_1)$, from which parts may be cut off by the box sides. It is of crucial importance that detunings on two sides of the curve (in the real space) have different signs.

Step 2. Check the resonance curve for integer points starting from a given integer point.
By construction, the curve has at least one integer point $(m_2, n_2)$. We calculate detunings for the point's neighbors in clockwise order:
\be
p_0=(m_2+1, n_2), p_1=(m_2+1, n_2+1), ... , p_7=(m_2+1, n_2-1)
\ee
and encounter one of the following possibilies:

2.0. We encounter a point with detuning zero, i.e. another resonant triad. We add it to the solution list and perform Step 2 for this point.

2.1. (The general case.) Detunings of two consequtive points $(p_i, p_{i+1}), i=1...7$ checked have different signs. (Here the index $i$ lies in $Z_8$, so that the pair $p_7, p_0$ does not fall out of consideration.) There lies a point of the resonance curve on the interval $[p_i, p_{i+1}]$. One coordinate of this point is the common coordinate of the pair (every pair has one coordinate in common, of course) and the other coordinate is not integer.

2.2. Some neighbor points fall outside the $L$-box.
Follow the border of the box until the next sign change
	(wchich signals the retun of the curve into the box), then
	follow the curve into the box in the manner described above.

These steps are iterated until we reach a point already visited before.
For each vector $\{m_1, n_1\}$ its resonant curve is treated in
$\mathcal{O}(L)$ steps, which amounts to $\mathcal{O}(L^3)$ overall
computation time. Moreover, the point couples constructed above, between
which the detuning function changes its sign, produce a set of quasi-resonances
much larger than just the immediate neighbors of exact resonant triads usually used.


In terms of resonant curves, quasi-resonances with detuning values under some limit $\e$ are integer box points belonging to some vicinity of the curve.
Let us introduce the notion of detuning surface, considering detuning as function of $(m_2, n_2)$ for fixed $(m_1, n_1)$ with $(m_2, n_2) \in \mathbb{R}$.
The algorithm outlined above actually does somewhat more than just find all integer solutions of the equations in question: for every resonant curve it also lists all integer points of the box that are its $1$-neighbors. A subset of them represents quasi-resonances. If this subset is not empty, we should check their immediate neighbors on the grid, and if some of them suffice quasi-resonance conditions, check their neighbors etc., until at some step no new quasi-resonance points are found. Simple properties of resonance curves and detuning surfaces allow us to prove that this search is exhaustive, i.e. \emph{all} quasi-resonances will be found through this procedure.
Moreover, as for each new quasi-resonance point found at some intermediate step we check only $\mathcal{O}(1) \le 8$ its neighbors, computational complexity is linear with respect to the number of quasi-resonances.

\section{Concluding remarks}\label{s:diss}
In this paper we have studied exact and quasi-resonances of two-dimensional Rossby/drift waves on the $b$-plane with periodic boundary conditions. This problem has been recently investigated by two group of researchers and the results obtained proved to be in a drastic contradiction.

Theoretical investigation presented in \cite{YaYo13} in the form of a strict mathematical theorem results in the statement that for big enough $\b$ flow dynamics is governed exclusively by resonant interactions, while quasi-resonances can be omitted.   On the other hand, numerical study of exact and quasi-resonances of two-dimensional Rossby/drift waves on the $b$-plane with periodic boundary conditions performed on the base of "complete classification of discrete resonant Rossby/drift wave triads", \cite{BH13},  states something very different. Namely, the resonance clustering is sparse, most of resonance clusters consist of one isolated triad or two connected triads, and the role of quasi-resonances in the energy transfer is crucial.

Being thus motivated to resolve this mystery, we performed a new study of resonance clustering for this case. Our results can be briefly formulated as follows.

We performed comparative study of the data sets (resonance triads,  \cite{BH13}, and resonant modes, \cite{YaYo13}) kindly provided by the authors of both papers. Comparison have been performed in the spectral domain $|m|,|n|\le 200$ while corresponding  resonant clustering is presented in the Fig.2, \cite{BH13}.  It turned out that algorithm developed in \cite{BH13} for computing exact resonances  produces only a small part of all solutions - about $15\%$ of all solutions in the chosen domain - and gives completely wrong resonance clustering. Instead of sparse clustering consisting mostly of isolated triads, we have established  that no isolated resonant triads exist for this case.  The smallest resonance cluster consists of 6 triads connected in a complicated way (shown in Fig.\ref{f:6tr}). This sixtuple of triads plays the role of a basic building block for resonance clustering in this case which is quite different from the other three wave systems studied previously, where indeed isolated triads play this role, \cite{CUP}. We assume that specifics of this resonance clustering yields a special mechanism of energy confinement by clusters of exact resonant modes which lacks in the other three wave systems presently known and even in the system of drift waves with other boundary conditions. Discussion of possible physical implications and analytical study of the dynamical system for a sixtuple of triads is forthcoming.

We have also established that an algorithm for computing quasi-resonances presented in \cite{BH13} is grossly incomplete and statistically biased while based on the incorrect assumption that  detuning value grows monotonously as the box size declines.  This algorithm does not only reduce the number of near-resonant triads constructed but also introduces a systematical bias - triads with smaller wave numbers are ignored more often than those with larger ones - rendering further considerations on them (both statistical and structural) virtually worthless.

We programmed this algorithm and checked this assumption for a few dozen arbitrary triads. We could not find a single triad for which this assumption holds, example of computation presented in Fig.\ref{f:detun}. Our software is free available for on-line computation,  \cite{link}. Another problem with this algorithm is that it looks for the triads with a small frequency detuning in \emph{a vicinity of an exact discrete resonant triad}.  However, as it follows from the Thue-Siegel-Roth theorem and has also been demonstrated for other wave systems, \cite{K07}, in this case the frequency detuning has a lower boundary and can not be arbitrary small. Accordingly, the smallest detuning found in \cite{BH13} in the vicinity of an exact resonance is of the order of $10^{-5}$.

However, one has to realize that the definition of a quasi-resonance (\ref{res-con-quasi})
does not require that one or two modes were part of some exact resonant triad. Full search in the domain $L=200$ renders detunings of the order of $10^{-11}$. In accordance with the Thue-Siegel-Roth theorem, the smallest detunings are found in the quasi-resonant triads whose distance to the nearest exact resonant is very big (the smallest detuning $\sim 8.95 \cdot 10^{-12}$, distance $\sim 21.7$, second best detuning $\sim 9\cdot10^{-12}$, distance $\sim 44$ etc.).

We  presented a detailed sketch of an algorithm for computing exact  resonances for periodic Rossby/drift waves on the $\b$-plane
based on the notion of resonant curves introduced in \cite{LongHigGill67}. This algorithm has cubic computational complexity, $T=\mathcal{O}(L^3)$, which results in quite realistic computation time (according to preliminary tests on a computer comparable to one used in \cite{BH13} estimated as a few hours to a few days, depending on implementation details) and has the evident advantage of finding \emph{all} exact solutions of the three wave kinematic resonant conditions. We also showed how a modification of this algorithm can be used to find all quasi-resonant solutions lying in a close vicinity of exact resonances.

It is a pity the authors of \cite{BH13} did not give details about the laptop they used
for their numerical studies. For some reason the authors performed full search only in the region $L=100$, considering full search in larger regions (including one of the main objects of the study - the $L=200$ domain) unfeasible.
For the sake of fair play we performed full search ("brute force") on the oldest laptop at hand and still running
(1.66 GHz single processor, 2 GB RAM, 32 bit system). The region $L=200$
was ready in $22$ minutes (the list of solutions coincides with data of M. Yamada and T. Yoneda, \cite{YaYo13}), and we see no reason to apply the heuristic algorithm developed in \cite{BH13},
losing some $84\%$ of exact solutions (695 from 828) and seven magnitudes of quasi-resonance detuning
$(10^{-5}$ versus $10^{-12}$) to this domain, or at least perform a comparison between the algorithm
proposed and full search.

\textbf{Remark}. Of course we performed full search with a few evident shortcuts,
like computing all $\o$-s and storing them in a $200 \times 200$ array, as well as taking
into account the symmetry group of order 12 mentioned above.

{\textbf{Acknowledgements.}} The authors are very much obliged to  M. D. Bustamante and U. Hayat, and to M. Yamada and T. Yoneda, for  supplying their data sets for our study.
AK  acknowledges support of AMS, Waidhofen an Ibbs, Austria. EK acknowledges  P. Diamond, \"{O}. Gurcan, A. Smolyakov and other participants of the
 7th Festival de The\`{'o}rie  "Reduced models of complex plasma dynamics" (8–26 July 2013,
Aix-en-Provence, France) for interesting discussions and most useful comments.
 This research has been supported by
the Austrian Science Foundation (FWF) under projects
P22943-N18 and and P24671.

\end{document}